\documentclass[review]{elsarticle}
\usepackage{amsmath,amsfonts,amsthm,bm}
\usepackage{natbib}
\renewcommand{\vec}[1]{\bm{#1}}
\usepackage{multirow}
\newtheorem{theorem}{Theorem}
\newtheorem{remark}{Remark}
\usepackage[english]{babel}
\usepackage[utf8]{inputenc}
\usepackage{graphicx}
\numberwithin{equation}{section}
\numberwithin{theorem}{section}
\numberwithin{remark}{section}
\usepackage{booktabs}
\usepackage{listings}
\usepackage[hyphens]{url}
\usepackage[colorlinks,linkcolor=black]{hyperref}
\usepackage{comment}
\hypersetup{breaklinks=true}
\usepackage{subfig}
\usepackage{lscape,float}
\usepackage{setspace}
\usepackage{listings}
\usepackage{comment}

\journal{XXX}
\makeatletter
\def\ps@pprintTitle{%
 \let\@oddhead\@empty
 \let\@evenhead\@empty
 \def\@oddfoot{\centerline{\thepage}}%
 \let\@evenfoot\@oddfoot}
\makeatother

\begin{document}

\begin{frontmatter}

\title{Assessing method agreement for paired repeated binary measurements administered by multiple raters}

%% or include affiliations in footnotes:
\author[mymainaddress]{Wei Wang}

\author[mymainaddress,Medaddress1]{Nan Lin\corref{mycorrespondingauthor}}
\ead{nlin@wustl.edu}
\author[Medaddress2]{Jordan D. Oberhaus}
\author[Medaddress2]{Michael S. Avidan}

\cortext[mycorrespondingauthor]{Corresponding author}

\address[mymainaddress]{\mbox{Department of Mathematics and Statistics, Washington University in St. Louis, St. Louis, USA }}
\address[Medaddress1]{Division of Biostatistics, Washington University School of Medicine, St. Louis,  USA}
\address[Medaddress2]{Department of Anesthesiology, Washington University School of Medicine, St. Louis,  USA }

\begin{spacing}{1}
\begin{abstract}
Method comparison studies are essential for development in medical and clinical fields. These studies often compare a cheaper, faster, or less invasive measuring method with a widely used one to see if they have sufficient agreement for interchangeable use. 
Moreover, unlike simply reading measurements from devices, e.g., reading body temperature from a thermometer,  the response measurement in many clinical and medical assessments is  impacted not only by the measuring device but also by the rater. For example, widespread inconsistencies are commonly observed among raters in   psychological  or cognitive assessment studies due to different characteristics such as rater training and experience, especially in large-scale assessment studies when many raters are employed.
This paper proposes a model-based approach to assess agreement  of two measuring methods for paired repeated binary measurements under the scenario where the  agreement between two measuring methods and the agreement among raters are required to be studied simultaneously. Based upon  the generalized linear mixed models (GLMM), the decision on the adequacy of  interchangeable use is made by testing the equality of fixed effects of methods. Approaches for assessing method agreement, such as the Bland-Altman diagram and  Cohen's kappa, are also developed for repeated binary measurements based upon the latent variables in GLMMs. We assess our novel model-based approach by simulation studies and a real clinical application, in which patients are evaluated repeatedly for delirium with two validated screening methods. Both the simulation studies and the real data analyses demonstrate that our proposed approach can effectively assess  method agreement.
\end{abstract}
\end{spacing}

\end{frontmatter}

%\linenumbers

\section{Introduction}\label{sec:intro}
Method comparison studies are designed to compare two measuring methods used to measure the same quantity. One is typically a new measuring method, and the other is often an existing and widely used one.  The primary goal is to assess the extent of agreement between the two measuring methods and decide if they have sufficient agreement so that they can be used interchangeably. That is, it does not matter which method is being used to take the measurement as both give practically the same value. If two measuring methods agree well enough to be used interchangeably, we may prefer the one that is cheaper, faster, less invasive, or easier to use. This is the primary motivation behind method comparison studies.~\cite{choudhary2017} 
For example, in the diagnosis of delirium, the Confusion Assessment Method (CAM) \cite{inouye1990} is the most widely used diagnostic questionnaire battery, and the 3-Minute Diagnostic Interview for Confusion Assessment Method (3D-CAM) \cite{marcantonio2014} is a 3-minute delirium assessment based upon the CAM. The goal of the comparison is to assess the agreement between the time-consuming CAM and the time-efficient 3D-CAM, and hence determine whether they can be used interchangeably.

In a broad sense, the term `measuring method' may refer to a medical device, an instrument, a questionnaire battery or a human judge. In this paper, we use the  term `rater' specifically for human judges. In the medical context, a rater may refer to a doctor in a diagnostic process, or a clinical observer in a clinical trial. There are occasions when the response measurement does not necessarily require the rater to provide subjective assessment, e.g., when reading the patient's blood pressure from a monitor.  We use the term `recorder' specifically for the  human judge  whose subjective assessment is not required.  The `inter-rater reliability' is the term for agreement among raters, while the `method agreement' is the term for agreement between measuring methods.  Throughout the literature, reviews of assessing agreement  usually discuss method agreement and inter-rater reliability together in the sense that agreement indices can often be used for both method agreement and inter-rater reliability, though entities on which agreement is assessed are different.
Detailed review of  assessing agreement can be found in the literature.~\cite{balakrishnan2010, barnhart2007,  carstensen2011, lin2012} Major approaches are summarized below for both continuous and categorical measurements. 
\begin{itemize}
\item  Continuous measurements
\begin{itemize}
\item The limits of agreement (LOA) approach  introduced by Bland and Altman is a widely used technique for assessing agreement between  two measuring methods or two raters of interest.~\cite{altman1983, bland1986, bland1999, bland2007} The LOA approach is accompanied by the Bland-Altman diagram to visually show the difference between two measuring methods or two raters. 
\item The intraclass correlation coefficient (ICC)  is used for assessing
agreement  among multiple measuring methods or multiple raters based on ANOVA-type models.~\cite{bartko1976, eliasziw1994, mcgraw1996, muller1994, shrout1979} These models mainly differ in assuming random effects or fixed effects on measuring methods or raters. 
\item The concordance correlation coefficient~\cite{lin1989} (CCC) was originally developed to assess agreement between two measuring methods or two raters for paired measurements without replications.  The CCC modifies Pearson's correlation coefficient by additionally assessing how far the best fitting line of the data is from the 45-degree line through the origin, i.e., deviation from perfect agreement. It was later extended to multiple measuring methods or multiple raters for data with replications.~\cite{barnhart2005,lin2007} In general, the CCC reduces to the ICC under the ANOVA models used to define the ICC.~\cite{barnhart2007}

\end{itemize}
\item Categorical measurements
\begin{itemize}
\item The agreement  between two measuring methods or two raters can be measured by Cohen's kappa \cite{cohen1960} for binary data or  its variants, e.g., weighted kappa \cite{cohen1968} for ordinal data. Fleiss' kappa \cite{fleiss1971measuring} can deal with multiple measuring methods or multiple raters. These kappa statistics are popular chance-corrected measures of agreement for categorical data.
\item The technique of model-based analysis is also widely used in measuring agreement for categorical data.~\cite{carrasco2005b, gao2012, nelson2015, nelson2017} By assuming that the categorical data follow a generalized linear mixed model (GLMM), approaches for continuous measurements can be applied to assessing agreement for the observed data on the scale of a latent variable.  
\end{itemize}
\end{itemize}
 Barnhart et al.~\cite{barnhart2007} classified measures of agreement by  unscaled agreement indices and scaled agreement indices.  Measures of agreement are scaled agreement indices if the values have regularized magnitudes, e.g., the ICC, CCC and kappa statistics which all  range from $-1$ to 1, and otherwise are unscaled, e.g.,  LOA. 
 
The proposed methodology of assessing method agreement in this paper is motivated by a fairly common  situation  in psychological  or cognitive assessment studies when  the response measurement is affected by both the measuring method and the rater. 
For example, in the diagnosis of mental health disorder, neuropsychological tests are used to determine the presence of cognitive strengths and weaknesses that may be the result of  a psychological disorder.  There are a wealth of test batteries which combine a range of neuropsychological tests to provide an overview of cognitive skills on patients, e.g., the Neurobehavioral Cognitive Status Examination (NCSE) \cite{kiernan1987neurobehavioral} and the Mini-Mental State Examination (MMSE) \cite{tombaugh1992mini}.  These tests are usually designed questionnaires or interviews administered by neuropsychologists, so test scores are finally determined  by neuropsychologists. Consequently, the test score is simultaneously affected by both the test battery and the neuropsychologist. This is  different from reading measurements from a medical device, e.g., reading body temperature from a thermometer or reading blood pressure from a monitor. When reading  body temperature from a thermometer, the rater is simply a recorder whose subjective assessment is not required, and hence could rarely impact the response measurement. However, in psychological  or cognitive assessment,  widespread inconsistencies are commonly observed among  raters due to various experience and training background, especially in large-scale assessment studies when plenty of raters are involved in administration. 
A special feature of method comparison studies in such settings is that method agreement and inter-rater reliability shall be investigated simultaneously.  To our knowledge, no literature has so far focused on assessing method agreement under the scenario where method agreement and inter-rater reliability are required to be studied simultaneously.
For example, in a recent review of the use and psychometric properties of the  NCSE \cite{shea2017review}, the study of comparing the  NCSE with the MMSE only discussed the sensitivity which evaluates the ability of a method to discern small changes  by a relative measure of precision, but no further discussion was carried out for assessing method agreement between these two test batteries, and neither included  considerations on rater's effect.

Our approach of  assessing method agreement   is  based on a generalized linear mixed model (GLMM) framework. 
 The usage of GLMMs is motivated by the need of addressing several challenges simultaneously: 1)  incorporating  the influence of raters into assessing method agreement; 2) categorical outcomes; 3)
 longitudinal data; 4) missing or unbalanced data. 
We will illustrate  our approach using a clinical research study~\cite{wildes2016protocol} in which surgical patients  are evaluated for post-operative delirium using two assessment tools, i.e., the CAM and the 3D-CAM. 
In this illustration, all the aforementioned challenges are encountered.  The GLMM allows modeling of  multiple sources of fixed and random effects, diverse response distributions and covariance structures, and thus is an analysis framework  that can accommodate all those complexities. 
Throughout this paper, we consider the situation where the rater's effect is assumed random, but the two measuring methods for comparison  are assumed fixed effects.  This is motivated by the intention of generalizing the method agreement results to a large population of raters because our primary goal is assessing method agreement. 
In the GLMM-based framework, we used hypothesis testing of equal fixed effects of two measuring methods  to determine whether the two methods agree. Furthermore, our proposed methodology provides straightforward graphical technique and summary statistics to measure the extent of method agreement for paired repeated binary measurements. Although the classic Bland-Altman diagram is generalized to assessing method agreement for data with multiple continuous measurements per subject \cite{bland2007}, it is still an open question how to plot the Bland-Altman diagram for  longitudinal binary measurements with correlation within the subject.  Gao et al.~\cite{gao2012} pointed out that the kappa statistic for analyzing repeated measurements is limited because it is  intended for data with a single observation made by each rater on each subject. To address these limitations, we propose novel ideas of plotting the Bland-Altman diagram and calculating Cohen's kappa for repeated binary measurements  based on the latent variables in the GLMMs. We prove that the newly developed diagram maintains the independence between the vertical-axis and horizontal-axis variables as the classic Bland-Altman diagram does.
In addition to assessing method agreement, we provide a way to simultaneously evaluate the inter-rater reliability by the intraclass correlation coefficient (ICC).  In summary, our new  methodology fills a gap in the current agreement literature to provide a flexible modeling approach to assess the method agreement and  inter-rater reliability simultaneously for paired repeated binary measurements. Our methodology is versatile  based on the GLMM framework in the sense that it can be extended to various data structures in more complicated cases.
Nelson et al.~\cite{nelson2015, nelson2017} also employed the GLMM framework to assess inter-rater reliability for ordinal ratings. Their approach incorporated the rater and patient characteristics that may impact inter-rater reliability,  but it is not intended for assessing method agreement. Roy et al.~\cite{roy2009, roy2015}  used   hypothesis testing to assess method agreement for repeated continuous measurements based on the linear mixed model (LMM). Like the clinical application of comparing imaging measurement devices therein~\cite{roy2015}, their approach is intended for the situation with the rater simply serving as a recorder. It  is different from the case  in this paper where  the response measurement requires the rater's subjective assessment.  
Gao et al.~\cite{gao2012}  provided a way to assess agreement between two measuring methods for paired repeated binary measurements impacted by a fixed set of raters as long as neither of the two methods is a gold standard. In their model, only the subject's effect is assumed random, while both the method's effect and  the rater's effect are assumed fixed, which is different from the case in this paper where we consider the rater's effect assumed random.  With the rater's effect assumed fixed, their approach  assessed method agreement  for each rater, and thus it is ideal for the case with a small set of raters.

The rest of  the paper is organized as follows. In Section 2, we introduce our approach to determine whether  two measuring methods agree by hypothesis testing based upon the GLMM.    Measures of method agreement and inter-rater reliability are further provided in this section. In Section 3, simulation studies demonstrate the performance of our approach. It is further illustrated in Section 4 using real data with simultaneous CAM and 3D-CAM assessments. The conclusion and discussion  are given in Section 5.

\section{Methodology}
\subsection{The GLMM-based framework}
Consider a study of method agreement between two measuring methods with $I$ subjects, $J$ raters and $T_i$ time points for subject $i = 1,\ldots,I$. The term `subject' is used here to refer to the entity on which the measurements are taken, e.g.,  a patient in a clinical trial.  Each subject is measured repeatedly over time. At each time point, a pair of  measurements are taken with both methods administered by two different raters.  The data are usually not balanced in the medical context because patients may not be measured  at every time point due to involuntary absence or rejection to continuing participation. 
We label the two methods as `1' and `2', and let  $(y_{ijt1},y_{ij't2})$ denote  the pair of  binary measurements  from  Methods 1 and 2 on a randomly selected subject $i$  at time point $t$ recorded at the same time by two  different raters $j$ and $j'$ randomly selected from a population of raters.  To model the binary measurements (either 0 or 1), we use a generalized linear mixed model (GLMM) with a probit link function, given by 
$$y_{ijtm} | \gamma_{i}, \alpha_{jm} \sim \mbox{Bernoulli}(\pi_{ijtm} ),$$  
and
\begin{equation*}
\pi_{ijtm} =\Phi( \mu_{ijtm}),
\end{equation*}
where  $m=1,2$,  $i = 1,2,\ldots,I$, $j = 1,2,\ldots,J$,  $t=1,\ldots,T_i$, and $\Phi(\cdot)$ is the cdf of the standard normal. The linear predictor $\mu_{ijtm}$ is given as 
\begin{equation}\label{mu}
 \mu_{ijtm}=\beta_m+g \left( x_t\right)+ \gamma_{i}+\alpha_{jm}.
\end{equation}
Terms in \eqref{mu} are assumed as follows.
\begin{itemize}
\item $\beta_m$ is the fixed effect of Method $m$, $m\in \{1,2\}$;
\item $g\left( x_t \right)$ is a regression  function that describes the dependence on a time-dependent covariate $x_t$. This allows to incorporate longitudinal 
 measurements on  subjects when  the mean response value changes over  time.
 \item $\gamma_{i}$ is the random effect of subjects, and $\gamma_{i} \overset{iid} \sim N(0,\sigma^2_{\gamma })$;
\item $\alpha_{jm}$ is the random effect of raters within Method $m$, and $\alpha_{jm} \overset{iid} \sim N(0,\sigma^2_{\alpha m})$.
\end{itemize}
Notice the dependence of the variance components on $m$, which means that we allow them to vary across different methods.

The probit link function $\Phi(\cdot)$ allows assessing method agreement and inter-rater reliability  on the latent scale.  Define the latent variable
$\tilde y_{ijtm}$ as 
\begin{equation}\label{latent}
\tilde y_{ijtm}=\mu_{ijtm}+\tilde \varepsilon_{ijtm},
\end{equation}
where  $\tilde \varepsilon_{ijtm}   \sim N(0,1)$ denotes the random error on the latent scale. The relationship between  the latent variable $\tilde y_{ijtm}$ and the observed $y_{ijtm}$ is 
 \[ y_{ijtm}=
\begin{cases} 
     1, & \ \mbox{if } \tilde y_{ijtm}>0;\\
       0 , & \mbox{ otherwise.}
\end{cases}\]   

Since we consider measurements on each subject  collected over time, we allow dependence in the within-subject errors of each method while letting the errors arising from different methods or from different subjects remain independent. That is, we assume that 
\begin{equation*}\label{cov}
Corr(\tilde \varepsilon_{ijtm},\tilde \varepsilon_{ij't'm'})=
\begin{cases} 
h\left(|t-t'|, \vec w\right), & \ \mbox{if } m=m';\\
 0 , & \mbox{if } m \ne m',
\end{cases}
\end{equation*}
where $h$ is a specified correlation function, and $\vec w$ is a vector of covariance parameters. For example, if the correlation matrix is AR(1) structured, the correlation function is 
\begin{equation}\label{ar1}
h\left(|t-t'|, \rho \right)=\rho^{|t-t'|}.
\end{equation}

From \eqref{mu}, the difference in the  latent variable between Methods 1 and 2 for subject $i$ at time point $t$ is 
$$D_{it}:=\tilde y_{ijt1}-\tilde y_{ij't2}=(\beta_1-\beta_2)+(\alpha_{j1}-\alpha_{j'2})+(\tilde \varepsilon_{ijt1}-\tilde \varepsilon_{ij't2}),$$
with marginal mean difference 
$$E(D_{it})=\beta_1-\beta_2.$$ Therefore, 
to decide whether the two methods   agree, it is equivalent  to testing  the following hypothesis,
\begin{equation}\label{H0}
H_0: \beta_1=\beta_2 \qquad vs. \qquad H_1: \beta_1 \ne \beta_2.
\end{equation}

If $H_0$ is rejected, we conclude that there is a significant difference between the two measuring methods. 
Otherwise,  we fail to reject the hypothesis that the two measuring methods agree, and then measures of method agreement introduced in the next section would further assess the extent of agreement.

%%%%%%%%%
\subsection{Measures of method agreement}
\subsubsection{Limits of agreement}
The limits of agreement (LOA) approach is popular for assessing agreement in the medical context.  Consider a set of $n$ subjects, each measured by both measuring methods, which results in $n$ pairs of measurements with one pair per subject. Assume that the underlying  difference between the two paired measurements  is $D$ and  $D \sim N(\xi, \nu^2)$. The LOA approach takes the interval $(\xi-1.96\nu, \xi+1.96\nu)$ covering the middle 95\% of the population of $D$ as the measures of agreement. The two  bounds of this  interval are estimated by $\hat \xi \pm 1.96\hat \nu$, which are called the 95\% limits of agreement. In the case of paired measurements data, the estimators $\hat \xi$ and $ \hat \nu^2$ are taken as the sample mean and sample variance of the observed differences, respectively.
If the two limits of agreement fall within  pre-specified margins $\pm \delta$, one can conclude that  the two measuring methods have sufficient agreement for interchangeable use. For example, the pre-specified margins $\pm \delta$ may refer to clinically acceptable difference in the medical context. Although $\delta$ is recommended to be specified in advance, it is rarely done so in practice. Instead, agreement of methods is often evaluated by judging whether the bounds of the interval $(\hat \xi - 1.96\hat \nu,\hat \xi + 1.96\hat \nu)$ are unacceptably large. \cite{choudhary2017}

The Bland-Altman diagram, as part of the LOA approach, is a popular graphical tool to evaluate method agreement. It plots the difference between the two paired measurements on the vertical axis against the average of the two measurements  on the horizontal axis to display the data. Further, three horizontal lines are added: one for the mean difference $\hat \xi$,  and one each for the two limits of agreement $\hat \xi \pm 1.96\hat \nu$. When two methods agree, the points in the Bland-Altman diagram scatter around zero in a random manner, and 95\% of the differences are expected to lie within the two lines corresponding to the two limits of agreement.

Barnhart et al.~\cite{barnhart2016choice}  concluded that among the existing agreement indices good for both continuous and categorical data, the coverage probability (CP) is the preferred agreement index on the basis of its consistent evaluation of data quality across multiple reviewers, populations, and continuous/categorical data. However, the classic Bland-Altman diagram is used for only continuous data. Compared to other agreement indices, the most appealing feature of the Bland-Altman diagram is that it visually shows the difference between two measurements, which is friendly to non-statisticians in the medical and clinical fields. Although the  Bland-Altman diagram is later generalized to assessing method agreement for data with multiple continuous measurements per subject \cite{bland2007}, the implicit assumption for the model therein is that there is no time effect and correlation in the difference between paired measurements. Thus, it is still an open question how to plot the Bland-Altman diagram for longitudinal binary measurements. In other words, there may be time effect and correlation among repeated measurements within the same subject.
 We shall show our novel idea of  plotting the Bland-Altman diagram with repeated measurements over time in our model-based setting. We aim to have one point per subject as the individual-level summary measure in the diagram, and also maintain the independence between the vertical-axis and horizontal-axis variables as the classic Bland-Altman diagram does.

One natural idea of the individual-level summary measure for each method is 
\begin{equation}\label{target}
\beta_m+\frac{1}{T_i}\sum^{T_i}_{t=1} g(x_t) +\gamma_{i}
\end{equation}
which averages  fixed time effect over time points and removes rater's effect. From the Bayesian perspective, we have the prior information that  for any rater $j \in \{ 1,\ldots,J\}$,
$$\mu_{ijtm}\Big \vert \beta_m+\frac{1}{T_i}\sum^{T_i}_{t=1} g(x_t) +\gamma_i \overset{iid}\sim N\left(\beta_m+\frac{1}{T_i}\sum^{T_i}_{t=1} g(x_t) +\gamma_{i},\sigma_{\alpha m}^2\right).$$ Then, the posterior distribution of \eqref{target} is given by the normal distribution
\begin{equation}\label{plotdiff}
N \left(\overline {\mu}_{im}, \left( \frac{1}{\sigma^2_{\gamma}} +\frac{J}{ \sigma^2_{\alpha m}}\right)^{-1} \right),
\end{equation}
where 
\begin{equation}\label{mu_im}
\overline {\mu}_{im}=  \beta_m + \frac{1}{T_i}\sum^{T_i}_{t=1} g(x_t) +\frac{J \sigma_{\gamma}^2}{J\sigma_{\gamma}^2+\sigma_{\alpha m}^2}  \left( \gamma_{i} +\frac{1}{J} \sum_{j=1}^J \alpha_{jm}\right), \ \  m=1,2.
\end{equation}
See Appendix A for detailed derivation of the posterior distribution \eqref{plotdiff}.

Note that, when   $J$  is large, the variance of the above normal distribution in  \eqref{plotdiff} goes to 0. The quantity  $\beta_m+ \frac{1}{T_i}\sum^{T_i}_{t=1} g(x_t)+\gamma_{i}$  on subject $i$ is then  measured by $\overline {\mu}_{im}$ accurately. 
Therefore, in order to measure the difference between two measuring methods for subject $i$ after eliminating the rater's effect, one natural idea is to use  the quantity
$\overline {\mu}_{i1}-\overline {\mu}_{i2}$ when  the number of raters, $J$,  is large.

\begin{remark}
In practice, when a dataset consists of only a few raters, we could consider another scenario with  a large value of $J \sigma^2_{\gamma}/\sigma^2_{\alpha m}$. If  the rater's variance $\sigma^2_{\alpha m}$ is relatively small, the value of $J \sigma^2_{\gamma}/\sigma^2_{\alpha m}$ could still be large even with only a few raters in the dataset. A relatively small value of rater's variance usually implies good agreement among raters.  For example, in the medical context, training is required to guarantee that doctors give consistent diagnoses for the same patient. In this case, in spite of only a few raters, the quantity $\overline {\mu}_{i1}-\overline {\mu}_{i2}$ is still proper to measure the difference between two measuring methods for each subject. Simulation results for this scenario are given in the supplementary materials.
\end{remark}

In practice,  $\overline \mu_{im}$ in \eqref{mu_im} can be evaluated based on 
the empirical best linear unbiased predictor~\cite{jiang2017asymptotic} (EBLUP) $\hat { \mu}_{im}$. The EBLUPs can be  provided by PROC GLIMMIX in SAS 9.4\cite{sas2017sas}. The following Theorem \ref{thm: equal} indicates that we can plot  a Bland-Altman diagram based on  the paired EBLUPs $\left( \hat \mu_{i1}, \hat \mu_{i2}\right)$  for $i =1,2,\ldots,I$,  on the  latent scale to visually show the difference between two methods. It is worth noting  that each subject is represented by one point in the Bland-Altman diagram.
  
\begin{theorem} \label{thm: equal}
Suppose that the fixed effects of methods satisfy $\beta_1 =\beta_2$. If the number of raters $J \to \infty$,  the points  $((\hat {\mu}_{i1}+\hat {\mu}_{i2})/2, \hat {\mu}_{i1}-\hat {\mu}_{i2})$ in the Bland-Altman diagram based on  the EBLUPs $\hat \mu_{i1}$ and $ \hat \mu_{i2}$  for all patients  $i =1,2,\ldots,I$,  scatter around zero in a random manner. That is,  for any $i \in \{1,\ldots,I\}$, as $J \to \infty$,
\begin{enumerate} 
\item [(a)]  $ \hat {\mu}_{i1} - \hat {\mu}_{i2}$ is uncorrelated with $(\hat {\mu}_{i1}+\hat {\mu}_{i2})/2$;
\item [(b)] $E( \hat {\mu}_{i1})-E(\hat {\mu}_{i2})\to 0$.
\end{enumerate}
\end{theorem}

\bigskip
In practice,  we can  further plot paired measurements  $\left( \Phi\left(  \hat \mu_{i1}\right) ,\Phi\left(  \hat \mu_{i2}\right) \right)$ to obtain the Bland-Altman diagram on the probability scale. If the distribution of the values on the probability scale is severely skewed, we can do transformation and then plot the Bland-Altman diagram of the transformed data 
\begin{equation}\label{log}
\left( \log\{\Phi\left(  \hat \mu_{i1}\right)\} ,\log\{\Phi\left(  \hat \mu_{i2}\right) \}\right).
\end{equation}

Alternatively, Cohen's kappa introduced in the following section  could also provide a way to measure the extent of method agreement.

\subsubsection{Cohen's kappa}\label{sec:kappa}
We have so far used the unscaled agreement index LOA to evaluate the method agreement based upon the fitted GLMM. In this section, we will show how to calculate the scaled agreement index Cohen's kappa based upon the fitted GLMM. For each $i\in \{1,2,\ldots,I\}$ and $m \in \{1,2\}$, the predicted $0$-$1$ binary score of the $i$th subject by Method $m$ is defined as
\[  \hat y_{im}=
\begin{cases} 
      1 , & \ \mbox{if } \hat \mu_{im} >0,\\
       0 , & \mbox{ otherwise.}
   \end{cases}\]
Here, $\hat \mu_{im}$ is the EBLUP that we used for the LOA approach in the previous section. 

Let $Y_1$ and $Y_2$ denote the outcomes $\hat y_{im}$ by Methods 1 and 2, respectively. The outcomes can then be presented as a  contingency table described in Table \ref{contingency}.

\begin{table}[h!]
\centering
\caption{Contingency table of outcomes.}\label{contingency}
\begin{tabular}{c| c c}
\hline
& $Y_1=0$ &$Y_1=1$\\
\hline
$Y_2=0$& $a$ & $b$\\
$Y_2=1$&  $c$&$d$\\
\hline
\end{tabular}
\end{table}
Here, `a', `b', `c' and `d' denote the numbers of patients with four possible combinations of outcomes measured by Methods 1 and 2. For example, 
`a' denotes the number of patients whose predicted outcomes from GLMM   are 0 from both measuring methods.

 Cohen's kappa $\kappa$ is given by 
\begin{equation}\label{kappa}
\kappa=\frac{p_o-p_e}{1-p_e},
\end{equation}
where $$p_o=\frac{a+d}{a+b+c+d}\quad  \mbox{ and } \quad p_e=\frac{a+b}{a+b+c+d}\cdot \frac{a+c}{a+b+c+d}+\frac{c+d}{a+b+c+d}\cdot \frac{b+d}{a+b+c+d}\ .$$ A confidence interval for Cohen's kappa is given based on the variance estimates discussed by Fleiss, Cohen, and Everitt\cite{fleiss1969large}. The calculation of Cohen's kappa together with its confidence interval is implemented using the \textsf{R} package `\texttt{psych}'\cite{psych19}.

A limitation of the original Cohen's kappa is that its magnitude is sensitive to the underlying disease prevalence, which is described as the prevalence effect.\cite{oleckno2008epidemiology} For example, older patients may experience an increased prevalence of some disease, and are more likely to be diagnosed as `Positive'. For rare diseases, subjects are more likely to be diagnosed as `Negative'. In either case, Cohen's kappa would decrease due to the imbalance between $a$ and $d$ in Table~\ref{contingency}.  It is an overcorrection for chance agreement. If one applies Cohen's kappa directly to repeated measurements, the imbalance between $a$ and $d$ may significantly increase due to the correlation among repeated measurements. For example, the prevalence effect for rare diseases would become severe if each subject is measured at a number of time points, i.e., large $T_i$, because the number of `Negative' $a$ may keep increasing while the number of `Positive' $d$ remains unchanged as more observations are collected for each subject over time.  It is worth noting that  our predicted binary score $\hat y_{im}$ is a summarized predicted binary value for each subject  based upon the latent variable model,  in the sense that each subject has only one binary value of $\hat y_{im}$ rather than  repeated binary measurements over time in the observed dataset. Therefore, our approach is able to correct the prevalence effect in repeated measurements.
Furthermore, our GLMM-based approach could incorporate subject's characteristics related to the underlying disease prevalence, e.g., patient's age, and hence could remove these effects to avoid overcorrection for chance agreement. 

%%%%%%%
\subsection{Measures of inter-rater reliability }
Our framework can also measure  the inter-rater reliability within each measuring method. Note that the effect of raters in  \eqref{mu} is assumed random. As mentioned in Section 1, the ICC is appropriate to deal with raters randomly selected from a large population. Therefore, agreement among raters within  the $m$th method ($m=1, 2$) can be measured by the ICC based upon the  latent variable model \eqref{latent}. The ICC within the $m$th method is given by
\begin{equation}\label{icc}
ICC_{m}=\frac{Cov(\tilde y_{ijtm},\tilde y_{ij'tm})}{\sqrt{Var(\tilde y_{ijtm})  Var(\tilde y_{ij'tm})}}=\frac{\sigma^2_{\gamma}+  1}{\sigma^2_{\gamma}+\sigma^2_{\alpha m}+1}, \qquad m=1,2.
\end{equation}
The estimated ICC is given by simply plugging in the estimators of variance components.

%%%%%%%%%%%%%

\section{Simulation Study}
In the simulation, we set  the numbers of subjects, raters and time points as $I=100$, $J=30$ and $T_i=5$ for any $i=1,\ldots,I$.   
We will first demonstrate  our approach in Section 3.1 based on one simulated dataset for each setup. We will also present the averages of parameter estimates  based on 1000 Monte Carlo replicates.
In Section 3.2, we will  compare our approach with approaches that do not account for the rater's effect. The GLMM fitting is  implemented by PROC GLIMMIX in SAS 9.4\cite{sas2017sas}.

\subsection{Illustrative examples}\label{Illustrative eg}
We shall consider two setups of \eqref{mu} and \eqref{ar1}, one for the case where two measuring methods agree, and the other for the case where they disagree. We assign true values of parameters in (\ref{mu}) and (\ref{ar1}) as follows. 
\begin{itemize}
\item \textsl{( Model 1 )}  The two measuring methods agree. \\
\text{  } Set $\beta_{1}=\beta_{2}=1.6$. Hence, $\beta_{1}-\beta_{2}=0$. \\ 
\textsl{( Model 2 )}  The two measuring methods disagree.\\
\text{  } Set $\beta_{1}=2.2$ and $\beta_{2}=1.6$. Hence, $\beta_{1}-\beta_{2}=0.6$.
\item We consider the following setup of variance components.\\
Set $\sigma^2_{\gamma}=0.8$, $\sigma^2_{\alpha 1}=0.2,\sigma^2_{\alpha 2}=0.4$, $\rho=0.1$.
\item Set $g(x_t)=-0.5x_t$, and $x_t=t$, for $t=1,\ldots, T_i$.
\end{itemize}

Sections \ref{3.1.1} gives the results of  our approach for one simulated dataset for each setup, particularly to demonstrate the Bland-Altman plot. Section \ref{3.1.2} further presents the performance of our proposed method  based on 1000 simulated datasets.

\subsubsection{Results for method agreement}\label{3.1.1}
Tables \ref{tab1}-\ref{tab2} present the results at significance level 0.05. 
It shows that the estimates of $\beta_1-\beta_2$ are $0.0276$  under Model 1;  0.5706 under Model 2, which are  close to the true values 0 under Model 1 and $0.6$ under Model 2, respectively. Under Model 1, we do not reject the null hypothesis $\beta_1=\beta_2$ with  $p$-value  $0.8739$ . Under Model 2, the $p$-value for testing (\ref{H0}) is $0.0010$, which shows a  significant difference between $\beta_1$ and $\beta_2$.  All these testing results agree with the underlying truth.

\begin{table}[http!]
\centering
\caption{ Estimation for the difference between methods, $\beta_1-\beta_2$. }
\begin{tabular}{ c  c c c c }
\hline
\hline
%\multicolumn{5}{c}{\textbf{Difference of method Least Square Means}  }     \\                                                                                                                                                                                                                                                     
Model  &{Estimate}  &  {SE} & {P-value} &{95\% CI}  \\
\hline
Model 1 &0.0276 & 0.1731   &0.8739  & ($- 0.3211$, 0.3764)     \\
  \hline
Model 2&$0.5706	 $   &  0.1627	    &0.0010 & (0.2437, 0.8975)       \\
 \hline                                                          
\end{tabular}
\label{tab1}
\end{table}

\begin{table}[htpp]
\centering
\caption{ Estimation for variance components. }
\begin{tabular}{ c c c c }
%\multicolumn{5}{c}{\textbf{Covariance Parameter Estimates} }        \\        
\hline
 \hline
Model & Parameter&Estimate&SE\\
\hline
 \multirow{ 5}{*}{Model 1} 
  &$\sigma^2_{\gamma}$               &    0.7578&               0.1511     \\
 &$\sigma^2_{\alpha 1}$           &                 0.1804	  &        0.0770           \\
&$\sigma^2_{\alpha 2}$             &                   0.4886	      &            0.1707                 \\
&$\rho$          &       0.0626   &  0.0410    \\
\hline
 \multirow{ 5}{*}{Model 2} &$\sigma^2_{\gamma}$               &   0.7179	& 0.1424       \\
 &$\sigma^2_{\alpha 1}$           &           0.1842	& 0.0835      \\
&$\sigma^2_{\alpha 2}$             &                   0.3756	 &0.1316      \\
&$\rho$          &   0.0686 &	0.0435 \\

\hline     
                                                     
\end{tabular}
\label{tab2}
\end{table}

\begin{figure}[t!]
  \centering
  \subfloat[Model 1 on the latent scale.]{\includegraphics[width=0.45\linewidth]{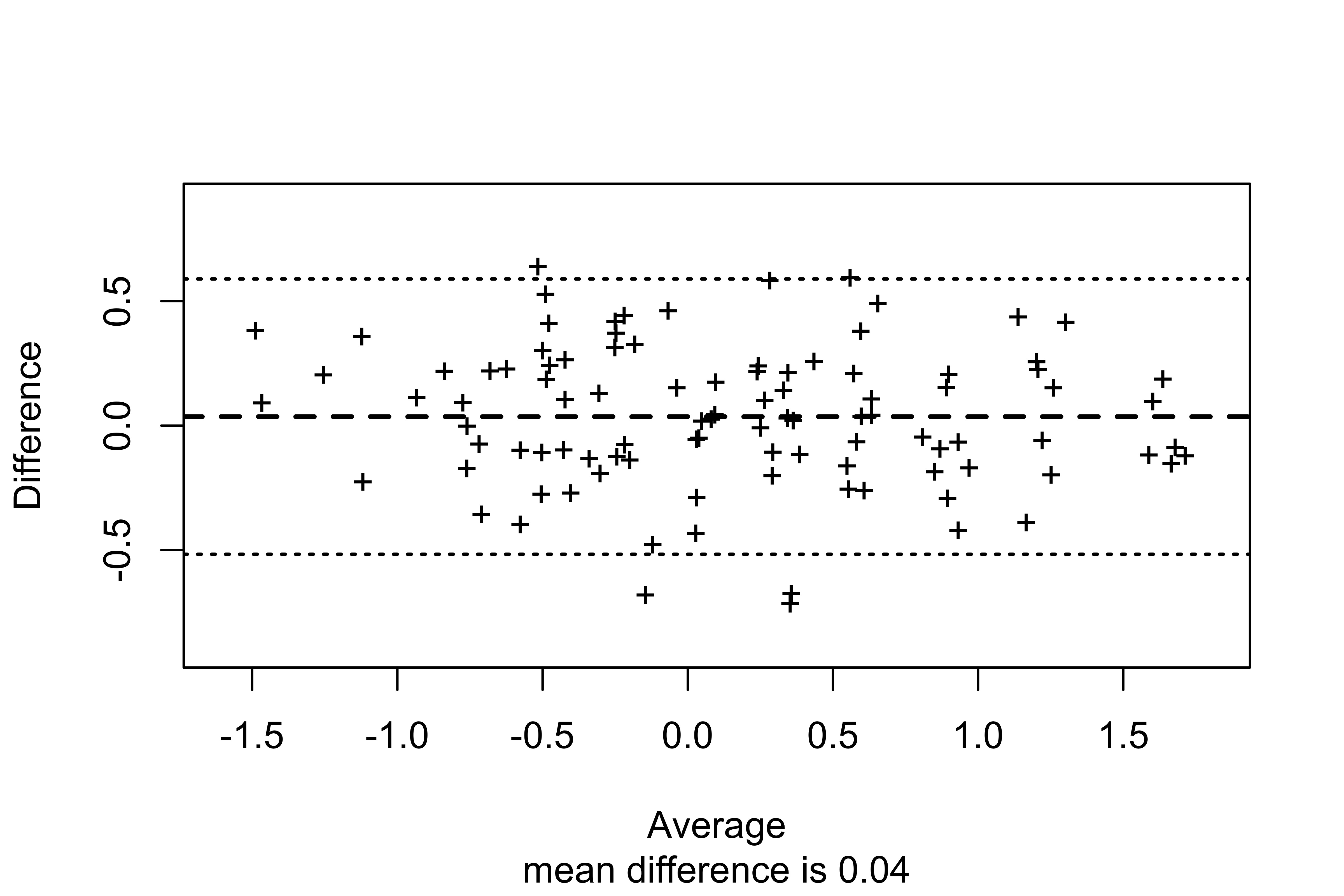}  \label{equalVarBA1}}
\qquad
  \subfloat[Model 1 on the probability  scale.]{\includegraphics[width=0.45\linewidth]{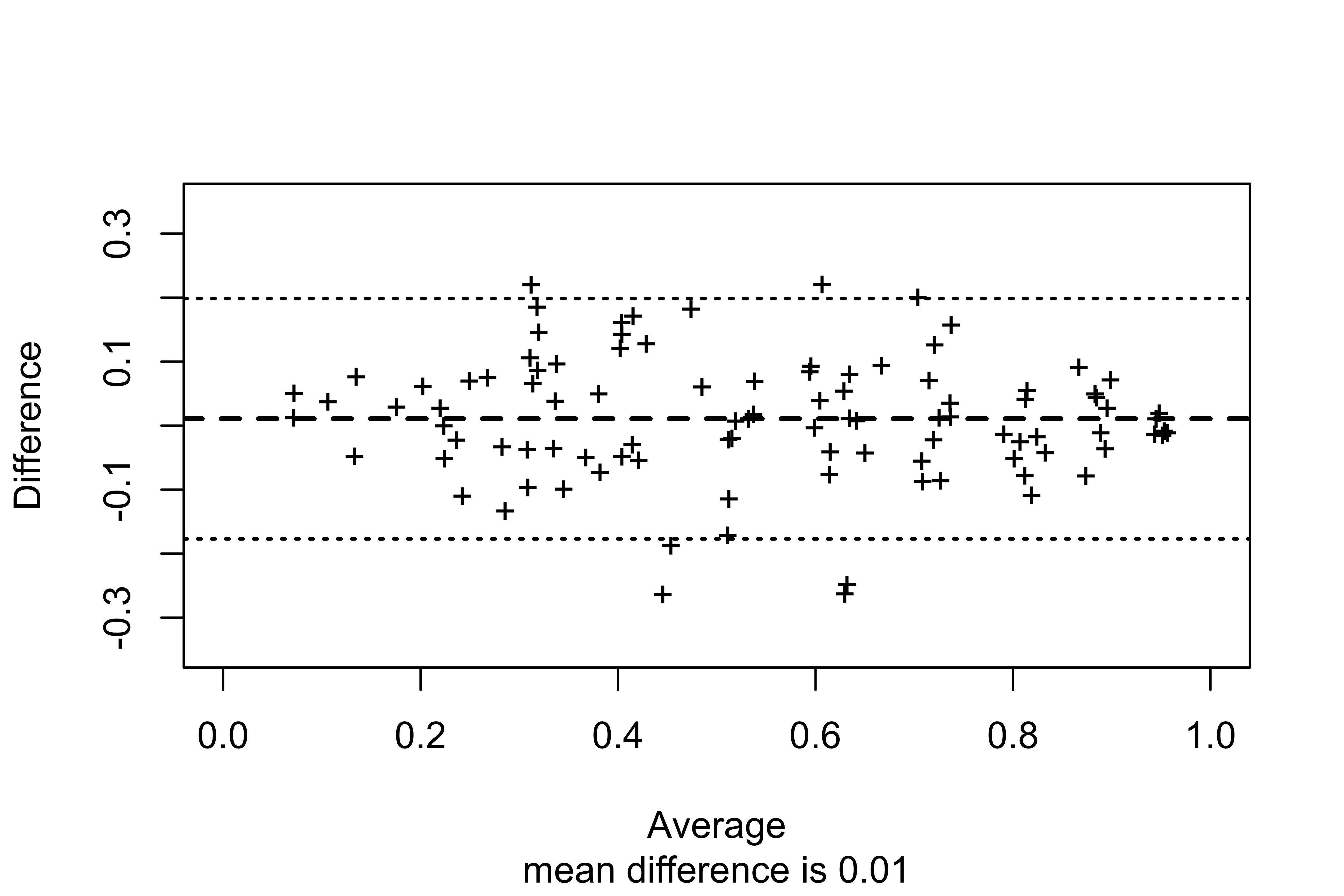}\label{equalVarBA2}}
 \\
 \subfloat[Model 2 on the latent scale.]  {\includegraphics[width=0.45\linewidth]{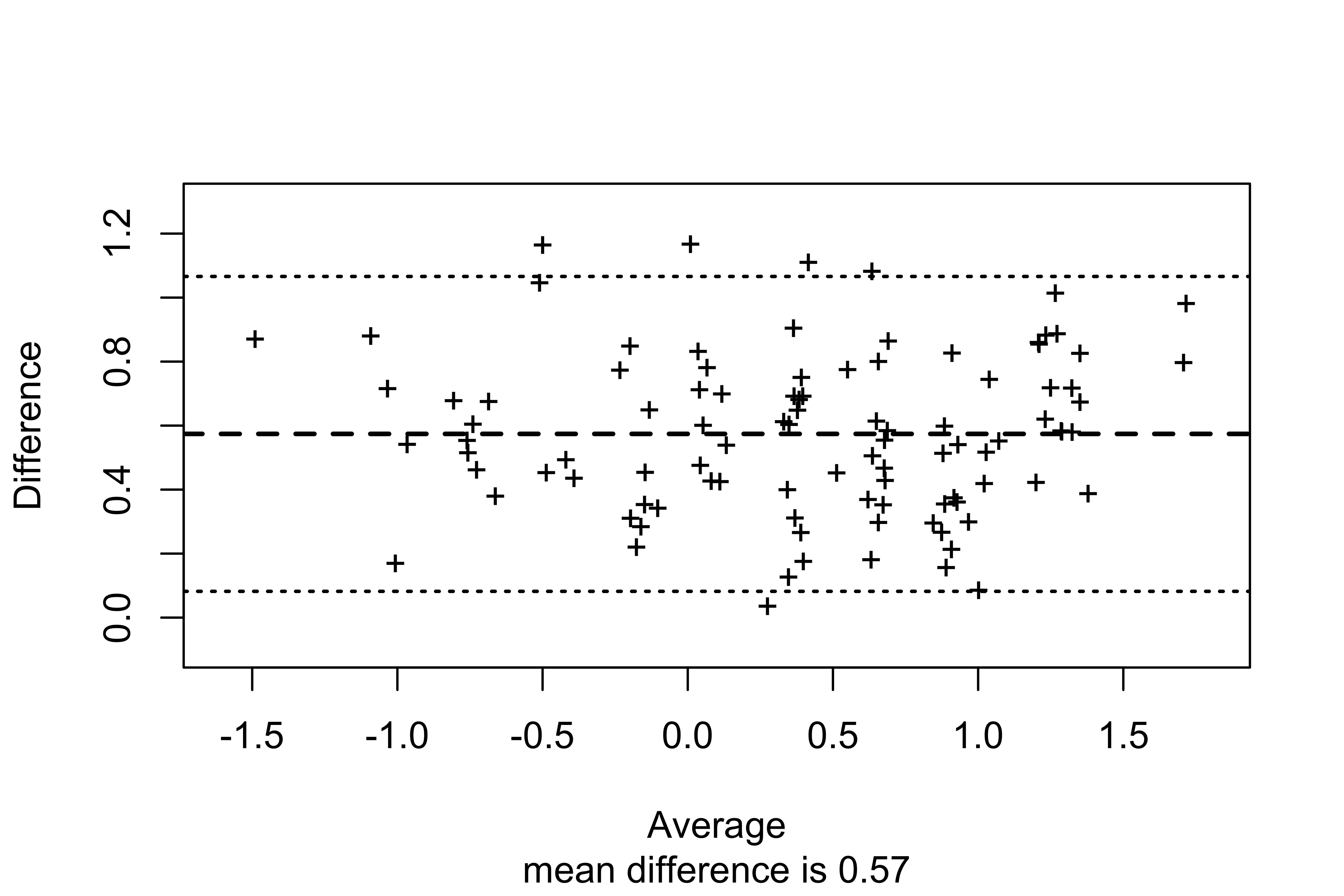}\label{equalVarBA3}}
  \qquad
 \subfloat[Model 2 on the probability scale.]  {\includegraphics[width=0.45\linewidth]{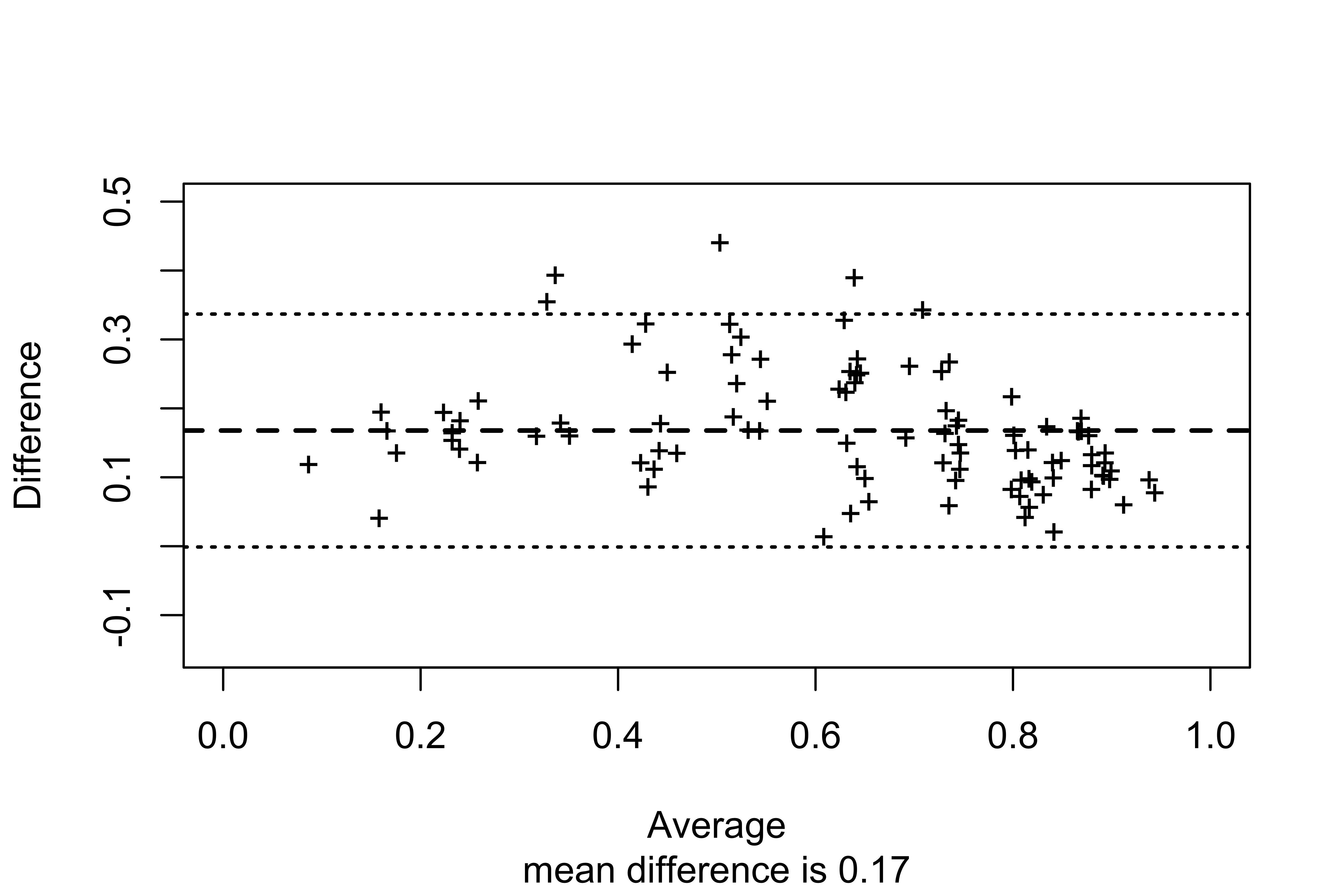}\label{equalVarBA4}}
  \caption{The Bland-Altman diagrams.}
\end{figure}

 The Bland-Altman diagrams on both the latent scale and the probability scale are shown in Figures \ref{equalVarBA1} - \ref{equalVarBA4}. 
As we can see in all  diagrams,   the points  are  around zero in a random manner,  so there is no systematic pattern. The dashed line shows the mean difference, and the dotted lines indicate the 95\% limits  of agreement. On both the latent scale and the probability scale, about 95 out of 100 points lie within the 95\% limits of agreement. Under Model 1, the mean difference is 0.04, which is close to the true difference 0. Under Model 2, the mean difference is 0.57, close to the true difference 0.6 and also nearly the same as the difference 0.5706 estimated by the GLMM shown in Table \ref{tab1}.

The ICCs for two measuring methods are calculated based on (\ref{icc}) and estimates of variance components from Table \ref{tab2}. Table \ref{tab3} summarizes the results of the ICCs. 
The true ICCs for Methods 1 and 2 are given by 0.9 and 0.8182, respectively. 
\begin{comment}
$${ICC}_1=\frac{  0.8   +1 }{   0.2+0.8    +1}=0.9,\quad \mbox{and} \quad {ICC}_2=\frac{   0.8  +1 }{   0.4+ 0.8   +1}=0.818.$$ 
\end{comment}
Therefore,  the simulated results in Table \ref{tab3} are  close to the true values of the ICCs for two measuring methods.

\begin{table}[htpp]
\centering
\caption{ The ICCs for inter-rater reliability. }
\begin{tabular}{ c c c c }
%\multicolumn{5}{c}{\textbf{Difference of method Least Square Means}  }     \\        
\hline \hline                                                                                                                                                                                                                                             
Model&{Method 1}  &  {Method 2} \\
\hline
Model 1 & 0.9069& 0.7825      \\
  \hline
{Model 2} &0.9032 &0.8206
    \\
\hline                                                          
\end{tabular}
\label{tab3}
\end{table}

We also  investigate the performance of our implementation of Cohen's kappa compared to naive Cohen's kappa, that is, directly applying Cohen's kappa to the observed data with correlated repeated measurements. For $\beta_1=\beta_2$, our approach gives a kappa value 0.82 with the 95\% confidence interval $(0.7, 0.93)$. On the other hand, for $\beta_1-\beta_2=0.6$,  our approach gives a kappa value 0.53 with the 95\% confidence interval $(0.36, 0.69)$. However,  the naive Cohen's kappas  have similar values in these two examples, with  values  0.3 and a  95\% confidence interval $(0.22, 0.38)$ for $\beta_1=\beta_2$, and  0.29 with a 95\% confidence interval $(0.21, 0.37)$ for $\beta_1-\beta_2=0.6$. Despite the change in the difference between $\beta_1$ and $\beta_2$, i.e., 0 vs 0.6, the two naive kappa values for observed data do not change much because  we find that the counts of disagreement $b+c$ are very similar due to correlated repeated measurements over time, i.e., both around 170 out of the total 500 observations. It is expected that the case with $\beta_1=\beta_2$ should show a higher agreement between two methods. Our approach gives a high kappa value 0.82, which suggests almost perfect agreement. However, naive  Cohen's kappa 
gives a very low value  0.3, which significantly deviates the underlying truth of method agreement.

\subsubsection{Parameter estimates}\label{3.1.2}
Table \ref{ave.est} presents averaged estimates of $\beta_1$, $\beta_2$ and ICCs over 1000 simulated datasets. It shows that the averaged estimates of ($\beta_1$, $\beta_2$) are all very close to the true values, i.e., $(1.6, 1.6)$ for Model 1, and $(2.2,1.6)$ for Model 2. 
The averaged ICCs are all very close to the true ICCs for Methods 1 and 2, i.e., 0.9 and 0.8182.

\begin{table}[htpp]
\centering
\caption{ Estimation for $\beta_1$, $\beta_2$ and ICCs. The results are based on 1000 replicates.}
\begin{tabular}{ c c c c }
%\multicolumn{5}{c}{\textbf{Covariance Parameter Estimates} }        \\        
\hline
 \hline
Scenario & Parameter&Estimate&SE\\
\hline
 \multirow{ 4}{*}{Model 1} 
&$\beta_1$               &    1.5659&              0.1968    \\
  &$\beta_2$               &    1.5697&             0.2087    \\
 &ICC$_1$           &                 0.8888	  &        0.0427          \\
&ICC$_2$             &                   0.8857	      &           0.0440                \\
\hline
 \multirow{ 4}{*}{Model 2} 
 &$\beta_1$               &   2.1657&              0.2189    \\
  &$\beta_2$               &    1.5943 &               0.2187     \\
 &ICC$_1$           &                 0.8161	  &        0.0565           \\
&ICC$_2$             &                   0.8102      &            0.0568            \\
\hline                                                        
\end{tabular}
\label{ave.est}
\end{table}

%%%%%%%%%%%%%%%%%%%%%%%
\subsection{Power comparison}
We compare our approach with approaches without taking the rater's effect into account, i.e.,
 \begin{equation}\label{noRater}
 \mu_{ijtm}=\beta_m+g \left( x_t\right)+ \gamma_{i}.
\end{equation}
The only difference is whether the linear predictor incorporates the term $\alpha_{jm}$ that accounts for  the random effect of raters.
We carry out a simulation study  to explore how the presence of unobserved rater heterogeneity would affect the size and power of test for method agreement in (\ref{H0}). We fix $\beta_2=1.6$ and let $\beta_1$ range from 1.6 to 2.8 by 0.1. Except for $\beta_1,\beta_2$, the setup for other parameters  is exactly the same as that in Section 3.1. We can explore the size and also how the power changes with the difference between $\beta_1$ and $\beta_2$.

\begin{figure}[http!]
\centering
\includegraphics[width=0.45\linewidth]{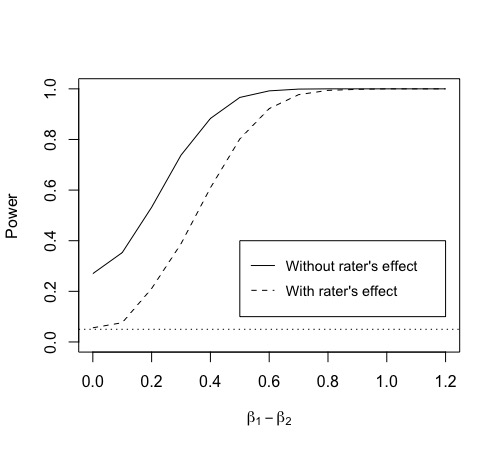}  
 \caption{Power comparison. The light grey dotted line indicates the significance level at 0.05.}\label{compower}
\end{figure}

Figure \ref{compower} shows  the power  of testing method agreement in (\ref{H0}) at significance level 0.05.  
The solid line represents the model   \eqref{noRater} not including the rater's effect, while the dashed line represents the  model  \eqref{mu} including the rater's effect. The simulation results are based on 1000 replicates at each value of $\beta_1$. The simulated sizes at significance level 0.05 are 0.056 and 0.270 for models fitted with and without rater's effect, respectively. Therefore, the model without rater's random effect cannot control the Type I error, though its power is higher than the power of the model with rater's random effect. 

%%%%%%%%%%%%%%%

%%%%%%%%%%%%%%%
\section{Real Data}
The Confusion Assessment Method (CAM)  \cite{inouye1990}  is the most widely used tool for delirium screening, but it is time consuming for clinical staff to administer on a routine basis. The 3-Minute Diagnostic Interview for Confusion Assessment Method (3D-CAM)~\cite{marcantonio2014} is a 3-minute delirium assessment based upon the CAM algorithm. The appeal of the 3D-CAM is that it takes less time than the CAM to screen for delirium. Therefore, it is meaningful to determine whether the agreement is sufficient between the two assessment tools so that CAM and 3D-CAM could be used interchangeably. In this section, we shall apply our approach of assessing method agreement to a real dataset for CAM and 3D-CAM comparison. 

Our data were collected from patients who underwent major elective surgery and were enrolled in the Electroencephalography Guidance of Anesthesia to Alleviate Geriatric Syndromes~\cite{wildes2016protocol}  (ENGAGES) trial, which looked at the effectiveness of electroencephalogram (EEG) guidance of anesthesia at preventing postoperative delirium at Barnes-Jewish Hospital in St. Louis, Missouri, USA. Patients enrolled in ENGAGES were 60 or older with at least a two-day hospital stay. We interviewed patients on post-operative days 0-5 using both the CAM and 3D-CAM at the same time but scored them independently in order to remain blind to the outcome of the other assessment. The dataset contains 42 pairs of readings (either `Positive' or `Negative') from 20 patients with 6 raters at 6 time points.  The data set is unbalanced since some patients were interviewed at only a couple of the time points.

Table $\ref{tab5}$ shows the result of testing the difference. Since the $p$-value is  0.523, the testing result indicates that the two methods may be used interchangeably.

\begin{table}[htpp]
\centering
\caption{ Comparison of the CAM and 3D-CAM. }
\begin{tabular}{ c c c c c }
%\multicolumn{5}{c}{\textbf{Difference of method Least Square Means}  }     \\     
\hline \hline                                                                                                                                                                                                                                             
 Difference & Estimate  & P-value & 95\% CI  \\
 \hline
$\beta_{CAM}-\beta_{3D\text{-}CAM}$ &-0.6446   &    0.5230	   & $(-3.7335, 2.4444)$ \\
\hline
 \end{tabular}
\label{tab5}
\end{table}

\begin{figure}[htpp]
  \centering
  \subfloat[The latent scale.]{\includegraphics[width=0.35\linewidth]{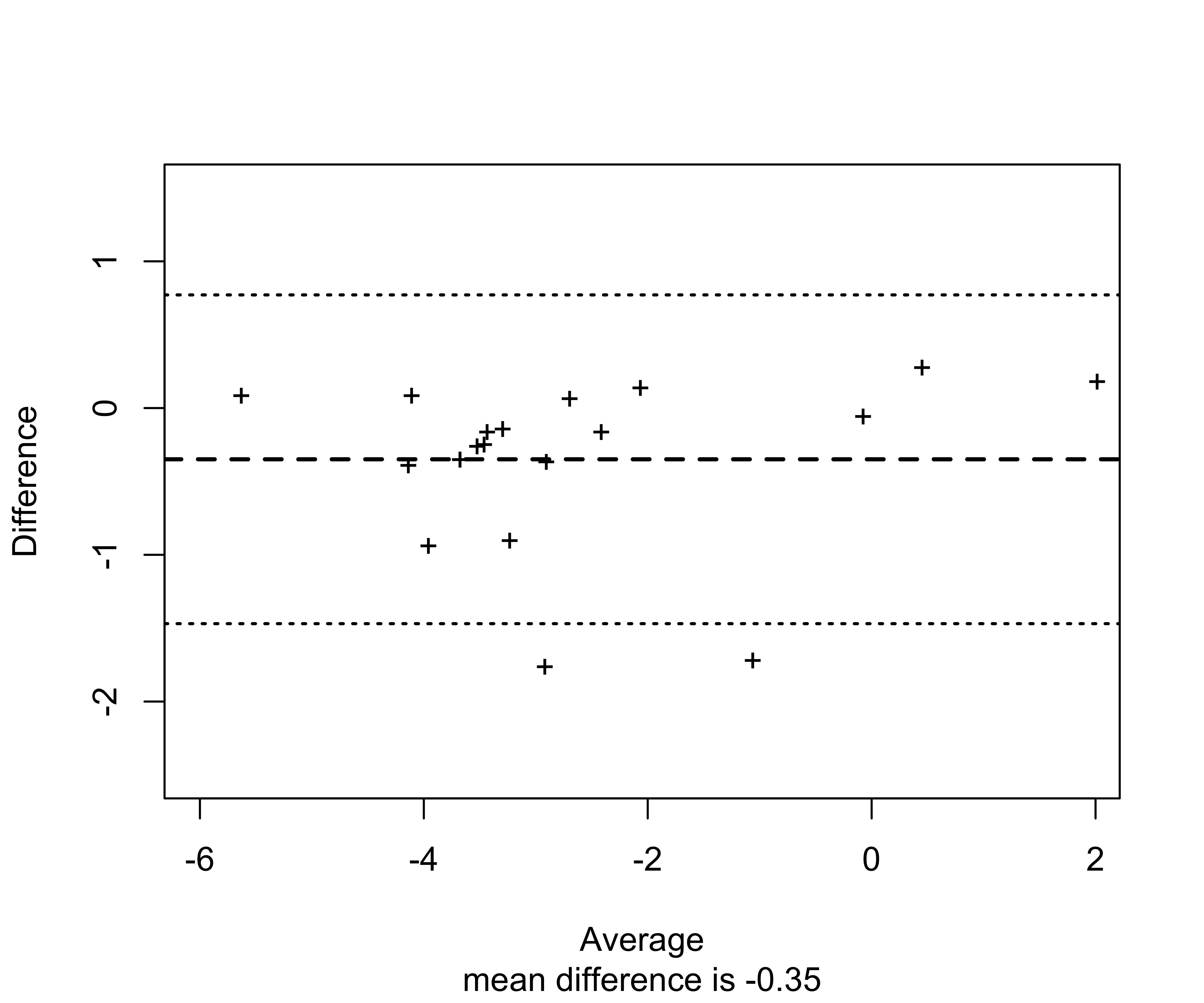}  \label{BA3}}
\qquad \quad
  \subfloat[The log-transformed probability scale.]{\includegraphics[width=0.35\linewidth]{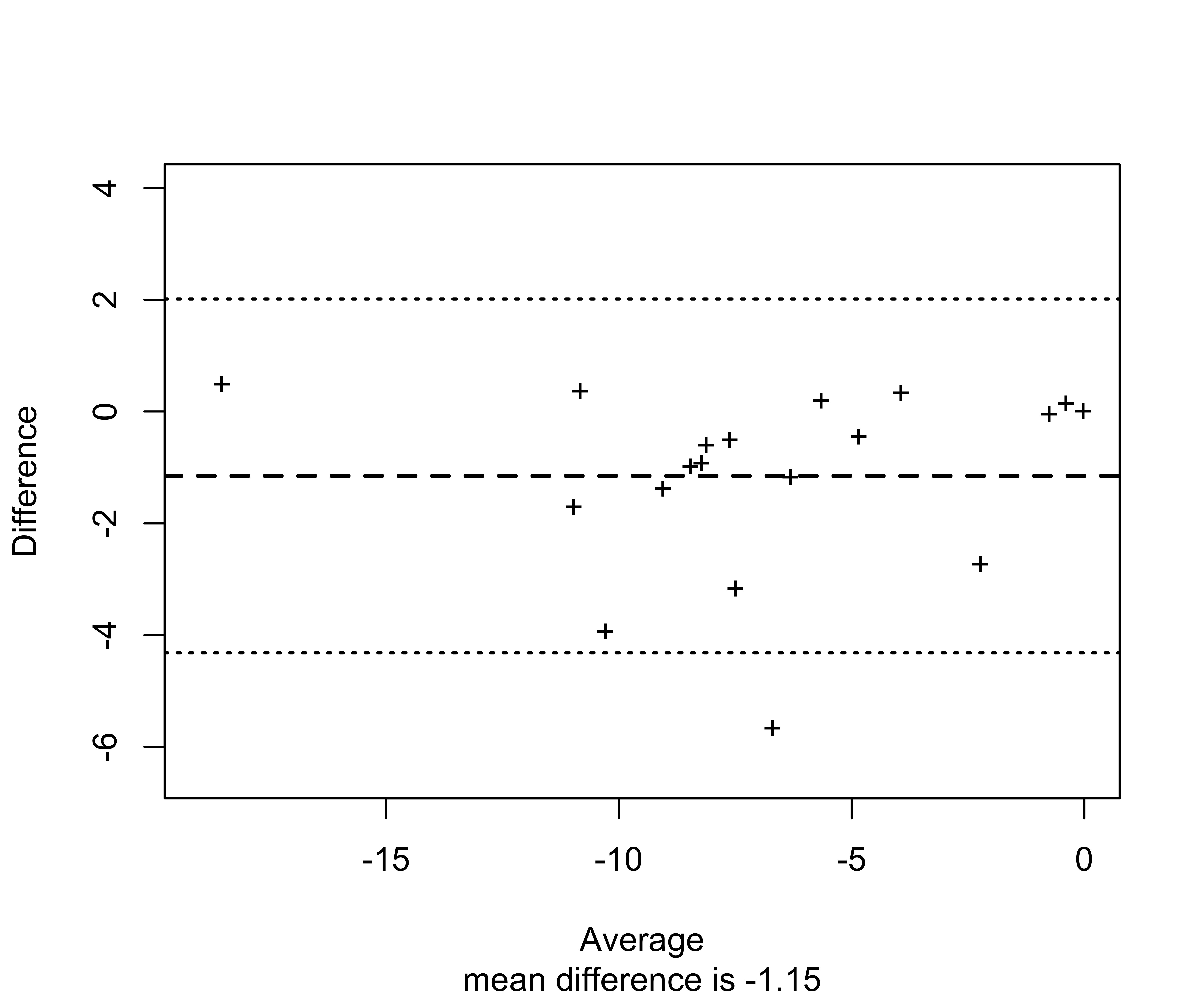}\label{BA4}}
  \caption{The Bland-Altman diagrams for real data.}
\end{figure}

The ICCs for the CAM and 3D-CAM are  0.99 and 0.85, respectively, which indicates that raters have an excellent degree of agreement for both measuring methods.
Next, we use the Bland-Altman diagram   to evaluate the degree of agreement. 
Figure \ref{BA3} shows the Bland-Altman diagram on the latent scale based upon 20 pairs of readings (one per patient), and Figure \ref{BA4} is drawn on  the log-transformed probability scale  \eqref{log}.

As we can see from Figure~\ref{BA3}, the mean difference on the latent scale is $-0.35$.  It might be hard for non-statisticians to interpret and set a pre-specified margin on the latent scale. Since the difference on the log-transformed scale $\log\{\Phi\left(  \hat \mu_{i1}\right)\} - \log\{\Phi\left(  \hat \mu_{i2}\right)\}= \log \{\Phi\left(\hat \mu_{i1} \right)/\Phi\left(\hat \mu_{i2} \right)\}$, the value obtained by taking the exponential of the mean difference would explain the ratio of the probability of scoring a `Positive'  between two methods, i.e., $\Phi\left(\hat \mu_{i1} \right)/\Phi\left(\hat \mu_{i2} \right)$. The mean difference in Figure 3(b) is $-1.15$, which indicates  the probability of `Positive'  measured by the  3D-CAM is on average $1/\exp\{-1.15\}=3.16$ times the probability by the CAM.
Although this point estimate on the log-transformed scale may show that  it is more likely to score `Positive'  by the 3D-CAM than by the CAM, the agreement of methods is evaluated by judging whether  the bounds of the 95\% confidence interval are unacceptably large on the log-transformed scale. 
The model-based Cohen's kappa calculated by \eqref{kappa} is exactly 1, which indicates perfect agreement between the CAM and 3D-CAM based on this dataset. In other words, the CAM and 3D-CAM are predicted to give the same values over these 20 patients, which supports the agreement between CAM and 3D-CAM. Our analysis result  is consistent with earlier studies in the literature
~\cite{ kuczmarska2016detection, vasunilashorn2016derivation}.
%%%%%%%%%%%%%%%

%%%%%%%%%%%%%%%
\section{Conclusion and discussion}
In this paper, we propose an approach for comparing two measuring methods with paired repeated binary data over time. Our GLMM-based framework incorporates both assessing method agreement and evaluating inter-rater reliability.
By treating methods as fixed effects, assessing method agreement is equivalent to testing the equality of fixed effects of methods.  Both simulation studies and applications to real data demonstrate the ability of our approach to make correct decisions on method agreement.

Provided that users are not satisfied with  simply a decision on whether the two methods agree or not, we further illustrate a novel way to implement the Bland-Altman diagram and Cohen's kappa on the latent variables based upon the GLMMs. Traditional scaled or unscaled agreement indices may provide misleading conclusions  with repeated data over time because of correlations among repeated measurements on the same subject, but our approach correctly measures the method agreement by accommodating the dependency in the GLMM.

Our methodology is versatile  based on the GLMM framework in the sense that it can be extended to various data structures in more complicated cases.
In general, the measurements could be continuous, binary or ordinal, while this paper focuses on repeated binary measurements.
The theory of our approach requires the assumption $J\to \infty$, which is a common condition for approaches based upon GLMMs to guarantee asymptotically consistent estimates of fixed effects and variance components for random effects. Although our approach theoretically intends to deal with large-scale studies in which there are  large numbers of raters and subjects, the additional simulation study in the supplementary materials shows that  our approach still performs well with only a few raters in the dataset as long as there is reasonably good agreement among raters.
Throughout this paper, we consider  the rater's effect as random by the intention of generalizing the method agreement results to a large population of raters. On the other hand, if the rater's effect is assumed fixed, there is a straightforward extension of the  hypothesis testing procedure based on the GLMM. Meanwhile, the newly developed Bland-Altman diagram and Cohen's kappa could be operated for each rater from a fixed set of raters.
 
%\nocite{*}% Show all bib entries - both cited and uncited; comment this line to view only cited bib entries;
\section*{Acknowledgments}
The ENGAGES study was funded by a National Institutes of Health grant to support pragmatic trials (1 UH2 HL125141, 5 UH3 AG050312). This study was also funded by the National Institutes of Health NIDUS Grant (NIA R24AG054259) and the Dr. Seymour and Rose T. Brown Endowed Chair at Washington University in St. Louis.

\appendix
\section{The posterior distribution of individual-level summary measure}
Let  ${\tau}_{ijm}$ be the average of $\mu_{ijtm}$ over time adjusted for  the fixed time effect $g(x_t)$,
 $$ {\tau}_{ijm}=\frac{1}{T_i}\sum^{T_i}_{t=1} \mu_{ijtm}-\frac{1}{T_i}\sum^{T_i}_{t=1} g(x_t). $$
 Then, $ {\tau}_{ijm}$ is modeled by the following mixed-effect model
 $$ {\tau}_{ijm}=\beta_m+\gamma_{i}+\alpha_{jm}.$$
 Notice that  for any rater $j \in \{ 1,\ldots,J\}$, we have 
  $$ {\tau}_{ijm}|\gamma_{i} \overset{iid} \sim N(\beta_m+\gamma_{i},\sigma_{\alpha m}^2),$$ and
 $$\beta_m+\gamma_{i} \sim N(\beta_m,\sigma_{\gamma}^2).$$
Let  ${\overline \tau}_{im}=\sum_{j=1}^J {\tau}_{ijm}/J$. Then 
$$\beta_m+\gamma_{i}| \overline{\tau}_{im} \sim N \left(\frac{ \sigma_{\alpha m}^2}{J \sigma_{\gamma}^2+\sigma_{\alpha m}^2 }\beta_m +\frac{J \sigma_{\gamma }^2}{J\sigma_{\gamma}^2+\sigma_{\alpha m}^2}   \overline{\tau}_{im}, \left( \frac{1}{\sigma^2_{\gamma}} +\frac{J}{ \sigma^2_{\alpha m}}\right)^{-1} \right),$$
so
\begin{equation*}
\beta_m+ \frac{1}{T_i}\sum^{T_i}_{t=1} g(x_t)+\gamma_{i}| \overline{\tau}_{im} \sim N \left(\overline {\mu}_{im}, \left( \frac{1}{\sigma^2_{\gamma}} +\frac{J}{ \sigma^2_{\alpha m}}\right)^{-1} \right),
\end{equation*}
where 
\begin{equation*}
\overline {\mu}_{im}=  \beta_m + \frac{1}{T_i}\sum^{T_i}_{t=1} g(x_t) +\frac{J \sigma_{\gamma}^2}{J\sigma_{\gamma}^2+\sigma_{\alpha m}^2}  \left( \gamma_{i} +\frac{1}{J} \sum_{j=1}^J \alpha_{jm}\right), \ \  m=1,2.
\end{equation*}

\section{Proof of Theorem 1}
\begin{proof}
From  \eqref{mu_im},  for $m\in\{1, 2\}$, the variance of $\overline {\mu}_{im}$ is
$$Var(\overline {\mu}_{im})=\frac{J\sigma^4_{\gamma}}{J\sigma^2_{\gamma}+\sigma^2_{\alpha m}} \to \sigma^2_{\gamma}, \mbox{  as } J\to \infty.$$
Then by checking 
\begin{equation*}\label{covdiff}
Cov(\overline {\mu}_{i1}- \overline {\mu}_{i2},\overline {\mu}_{i1}+\overline {\mu}_{i2})=Var(\overline {\mu}_{i1})-Var(\overline {\mu}_{i2})\to 0,
\end{equation*}
the difference $\overline {\mu}_{i1}-\overline {\mu}_{i2}$ is  uncorrelated with the average 
$(\overline {\mu}_{i1}+\overline {\mu}_{i2})/2$ as the number of raters $J \to \infty$. 

For $m \in \{1,2\}$, 
$$\overline \mu_{im} \overset{a.s.} \to \beta_m+ \frac{1}{T_i}\sum^{T_i}_{t=1} g(x_t)+\gamma_{i}, \qquad \mbox{as}  \quad J \to \infty.$$ 
Therefore, $(\overline {\mu}_{i1}   -\overline {\mu}_{i2} ) -( \beta_1-\beta_2) \overset{a.s.} \to 0$. If $\beta_1=\beta_2$, then $\overline {\mu}_{i1}   -\overline {\mu}_{i2} \overset{a.s.} \to 0 $. 
Similarly,
$(\hat {\mu}_{i1}   -\hat {\mu}_{i2}) -(\hat \beta_1-\hat \beta_2)  \overset{a.s.} \to 0$, where $\hat \beta_1, \hat \beta_2$ are EBLUEs of $\beta_1, \beta_2$. 
Since $\hat \beta_{m} - \beta_{m}  \overset{P} \to 0$,  we have $\hat  {\mu}_{i1}   -\hat {\mu}_{i2} \overset{P} \to 0$ if $\beta_1=\beta_2$.
By the normality assumption and assuming that $\beta_1=\beta_2$,
$$E(\hat \mu_{i1}-\overline {\mu}_{i1})-E(\hat \mu_{i2}-\overline {\mu}_{i2})=E(\hat \mu_{i1}-\hat \mu_{i2})-E(\overline {\mu}_{i1}-\overline {\mu}_{i2})\to 0, \quad \mbox{as} \quad J \to \infty.$$
By Theorem 13.2 in Jiang~\cite{jiang2010large} about the mean squared prediction error (MSPE) of the EBLUP, the difference between the MSPEs  $$E(\hat \mu_{i1}-\overline \mu_{i1})^2- E(\hat \mu_{i2}-\overline \mu_{i2})^2 \to 0,  \mbox{  as } J\to \infty.$$
Therefore, $$Var(\hat \mu_{i1}-\overline {\mu}_{i1})-Var(\hat \mu_{i2}-\overline {\mu}_{i2}) \to 0,  \mbox{  as } J\to \infty.$$ Then, the difference $(\hat {\mu}_{i1}-\overline {\mu}_{i1})-(\hat {\mu}_{i2}-\overline {\mu}_{i2})$ is also uncorrelated with 
 $(\hat {\mu}_{i1}-\overline {\mu}_{i1})/2+(\hat {\mu}_{i2}-\overline {\mu}_{i2})/2$ as the number of raters $J \to \infty$. 
 
Since $\overline {\mu}_{i1}-\overline {\mu}_{i2}$ is  uncorrelated with 
$\overline {\mu}_{i1}+\overline {\mu}_{i2}$, and $\overline {\mu}_{i1}   -\overline {\mu}_{i2} \overset{a.s.} \to 0 $,  $\hat {\mu}_{i1}   -\hat {\mu}_{i2} \overset{P} \to 0 $ as $J \to \infty$,  by  the fact that  
\[ \begin{split}
&\qquad Cov\left((\hat {\mu}_{i1}-\overline {\mu}_{i1})-(\hat {\mu}_{i2}-\overline {\mu}_{i2}), (\hat {\mu}_{i1}-\overline {\mu}_{i1})+(\hat {\mu}_{i2}-\overline {\mu}_{i2})\right)- Cov\left(\hat {\mu}_{i1}-\hat {\mu}_{i2},\hat {\mu}_{i1}+\hat {\mu}_{i2}\right)\\
& = Cov\left((\hat {\mu}_{i1}-\hat {\mu}_{i2})-(\overline {\mu}_{i1}-\overline {\mu}_{i2}), (\hat {\mu}_{i1}+\hat {\mu}_{i2})-(\overline {\mu}_{i1}+\overline {\mu}_{i2})\right)- Cov\left(\hat {\mu}_{i1}-\hat {\mu}_{i2},\hat {\mu}_{i1}+\hat {\mu}_{i2}\right)\\
&=Cov\left(\hat {\mu}_{i1}-\hat {\mu}_{i2},\hat {\mu}_{i1}+\hat {\mu}_{i2}\right)-2Cov(\hat {\mu}_{i1}, \overline {\mu}_{i1})+2Cov(\hat {\mu}_{i2}, \overline {\mu}_{i2})- Cov\left(\hat {\mu}_{i1}-\hat {\mu}_{i2},\hat {\mu}_{i1}+\hat {\mu}_{i2}\right)\\
&  \to 0 \mbox{,   \qquad  as } J \to \infty, 
\end{split}
\]
the difference between the EBLUPs, i.e.,  $\hat {\mu}_{i1}-\hat {\mu}_{i2}$, is also uncorrelated with $(\hat {\mu}_{i1}+\hat {\mu}_{i2})/2$, as $J \to \infty$.

Notice that  for $m=1,2,$ the mean of $\overline {\mu}_{im}$ is
$$E(\overline {\mu}_{im})= \beta_m + \frac{1}{T_i}\sum^{T_i}_{t=1} g(x_t) .$$
 If $\beta_1=\beta_2$,  then $E(\overline {\mu}_{i1})=E(\overline {\mu}_{i2})$. 
 As $E(\hat \mu_{i1}-\hat {\mu}_{i2})-E(\overline \mu_{i1}-\overline{\mu}_{i2})\to0$, $E(\hat  {\mu}_{i1})-E(\hat  {\mu}_{i2}) \to 0$  if  $\beta_1=\beta_2$.
 \end{proof}

\section{SAS code}
\begin{lstlisting}
/*For confidentiality, we cannot show the real data.*/ 
/*Instead, we provide the following pseudodata for user convenience.*/
data a;
 input  id $ time $ cam $ dcam $ rater_cam $ rater_dcam $;
 datalines;
 1 1 Negative Negative A B
 1 2 Negative Postive C D
 ... more data lines...
 ;
 
data b;set a;
y=cam; method='CAM';rater=rater_cam;id=id;time=time;time1=time;output;
y=dcam;method='3DCAM';rater=rater_dcam;id=id;time=time;time1=time;output;
keep y method rater id time time1;
 
proc glimmix data=b;
 class id time method rater;
 model y(event='Positive')= method time1/dist=binary link=probit ddfm=kr;
 random id;
 random rater/group=method;
 /*The correlation matrix of within-subject errors is AR(1) structured.*/
 random time/subject=id(method)  type=ar(1) rside; 
 /*Output linear predictor (p) and marginal linear predictor (np).*/
 output out=agreeout pred=p pred(noblup)=np;
 /*Test if the coefficients of method's effects are the same.*/
 lsmeans method/diff ilink cl;
run; 
\end{lstlisting}

\bigskip

\bibliography{MethodAgree2nd}

\begin{thebibliography}{10}
\providecommand \doibase [0]{http://dx.doi.org/}%

\bibitem{choudhary2017}
Choudhary PK, Nagaraja HN. {\it Measuring Agreement: Models, Methods, and
  Applications}.
\newblock New York, NY: John Wiley \& Sons; 2017.

\bibitem{inouye1990}
Inouye SK, {van Dyck} CH, Alessi CA, Balkin S, Siegal AP, Horwitz RI.
  Clarifying confusion: the confusion assessment method: A new method for
  detection of delirium. {\it Annals of Internal Medicine} 1990\string;
  113(12)\string: 941--948.

\bibitem{marcantonio2014}
Marcantonio ER, Ngo LH, O'connor M, et al. {3D-CAM}: derivation and validation
  of a 3-minute diagnostic interview for {CAM}-defined delirium: A
  cross-sectional diagnostic test study. {\it Annals of Internal Medicine}
  2014\string; 161(8)\string: 554--561.

\bibitem{balakrishnan2010}
Balakrishnan N. {\it Methods and Applications of Statistics in the Life and
  Health Sciences}.
\newblock New York, NY: John Wiley \& Sons; 2010.

\bibitem{barnhart2007}
Barnhart HX, Haber MJ, Lin LI. An overview on assessing agreement with
  continuous measurements. {\it Journal of Biopharmaceutical Statistics}
  2007\string; 17(4)\string: 529--569.

\bibitem{carstensen2011}
Carstensen B. {\it Comparing Clinical Measurement Methods: A Practical Guide}.
\newblock New York, NY: John Wiley \& Sons; 2011.

\bibitem{lin2012}
Lin L, Hedayat A, Wu W. {\it Statistical Tools for Measuring Agreement}.
\newblock New York, NY: Springer Science \& Business Media; 2012.

\bibitem{altman1983}
Altman DG, Bland JM. Measurement in medicine: The analysis of method comparison
  studies. {\it The Statistician} 1983\string; 32(3)\string: 307--317.

\bibitem{bland1986}
Bland JM, Altman D. Statistical methods for assessing agreement between two
  methods of clinical measurement. {\it The Lancet} 1986\string;
  327(8476)\string: 307--310.

\bibitem{bland1999}
Bland JM, Altman DG. Measuring agreement in method comparison studies. {\it
  Statistical Methods in Medical Research} 1999\string; 8(2)\string: 135--160.

\bibitem{bland2007}
Bland JM, Altman DG. Agreement between methods of measurement with multiple
  observations per individual. {\it Journal of Biopharmaceutical Statistics}
  2007\string; 17(4)\string: 571--582.

\bibitem{bartko1976}
Bartko JJ. On various intraclass correlation reliability coefficients. {\it
  Psychological Bulletin} 1976\string; 83(5)\string: 762--765.

\bibitem{eliasziw1994}
Eliasziw M, Young SL, Woodbury MG, Fryday-Field K. Statistical methodology for
  the concurrent assessment of interrater and intrarater reliability: Using
  goniometric measurements as an example. {\it Physical Therapy} 1994\string;
  74(8)\string: 777--788.

\bibitem{mcgraw1996}
McGraw KO, Wong SP. Forming inferences about some intraclass correlation
  coefficients. {\it Psychological Methods} 1996\string; 1(1)\string: 30--46.

\bibitem{muller1994}
M{\"u}ller R, B{\"u}ttner P. A critical discussion of intraclass correlation
  coefficients. {\it Statistics in Medicine} 1994\string; 13(23-24)\string:
  2465--2476.

\bibitem{shrout1979}
Shrout PE, Fleiss JL. Intraclass correlations: Uses in assessing rater
  reliability. {\it Psychological Bulletin} 1979\string; 86(2)\string:
  420--428.

\bibitem{lin1989}
Lawrence I, Lin K. A concordance correlation coefficient to evaluate
  reproducibility. {\it Biometrics} 1989\string; 45(1)\string: 255--268.

\bibitem{barnhart2005}
Barnhart HX, Song J, Haber MJ. Assessing intra, inter and total agreement with
  replicated readings. {\it Statistics in Medicine}\string; 24(9)\string:
  1371--1384.

\bibitem{lin2007}
Lin L, Hedayat A, Wu W. A unified approach for assessing agreement for
  continuous and categorical data. {\it Journal of Biopharmaceutical
  Statistics} 2007\string; 17(4)\string: 629--652.

\bibitem{cohen1960}
Cohen J. A coefficient of agreement for nominal scales. {\it Educational and
  Psychological Measurement} 1960\string; 20(1)\string: 37--46.

\bibitem{cohen1968}
Cohen J. Weighted kappa: Nominal scale agreement provision for scaled
  disagreement or partial credit. {\it Psychological Bulletin} 1968\string;
  70(4)\string: 213--220.

\bibitem{fleiss1971measuring}
Fleiss JL. Measuring nominal scale agreement among many raters. {\it
  Psychological Bulletin} 1971\string; 76(5)\string: 378--382.

\bibitem{carrasco2005b}
Carrasco JL, Jover L. Concordance correlation coefficient applied to discrete
  data. {\it Statistics in Medicine} 2005\string; 24(24)\string: 4021--4034.

\bibitem{gao2012}
Gao J, Pan Y, Haber M. Assessment of observer agreement for matched repeated
  binary measurements. {\it Computational Statistics \& Data Analysis}
  2012\string; 56(5)\string: 1052--1060.

\bibitem{nelson2015}
Nelson KP, Edwards D. Measures of agreement between many raters for ordinal
  classifications. {\it Statistics in Medicine}\string; 34(23)\string:
  3116--3132.

\bibitem{nelson2017}
Nelson KP, Mitani AA, Edwards D. Assessing the influence of rater and subject
  characteristics on measures of agreement for ordinal ratings. {\it Statistics
  in Medicine} 2017\string; 36(20)\string: 3181--3199.

\bibitem{kiernan1987neurobehavioral}
Kiernan RJ, Mueller J, Langston JW, Van~Dyke C. The Neurobehavioral Cognitive
  Status Examination: A brief but differentiated approach to cognitive
  assessment. {\it Annals of Internal Medicine} 1987\string; 107(4)\string:
  481--485.

\bibitem{tombaugh1992mini}
Tombaugh TN, McIntyre NJ. The mini-mental state examination: A comprehensive
  review. {\it Journal of the American Geriatrics Society} 1992\string;
  40(9)\string: 922--935.

\bibitem{shea2017review}
Shea T, Kane C, Mickens M. A review of the use and psychometric properties of
  the Cognistat/Neurobehavioral Cognitive Status Examination in adults
  post--cerebrovascular accident. {\it Rehabilitation Psychology} 2017\string;
  62(2)\string: 221--222.

\bibitem{wildes2016protocol}
Wildes T, Winter A, Maybrier H, et al. Protocol for the Electroencephalography
  Guidance of Anesthesia to Alleviate Geriatric Syndromes (ENGAGES) study: A
  pragmatic, randomised clinical trial. {\it BMJ Open} 2016\string;
  6(6)\string: e011505.

\bibitem{roy2009}
Roy A. An application of linear mixed effects model to assess the agreement
  between two methods with replicated observations. {\it Journal of
  Biopharmaceutical Statistics} 2009\string; 19(1)\string: 150--173.

\bibitem{roy2015}
Roy A, Fuller CD, Rosenthal DI, Thomas~Jr CR. Comparison of measurement methods
  with a mixed effects procedure accounting for replicated evaluations ({COM 3
  PARE}): Method comparison algorithm implementation for head and neck {IGRT}
  positional verification. {\it BMC Medical Imaging} 2015\string; 15(1)\string:
  35--45.

\bibitem{barnhart2016choice}
Barnhart HX, Yow E, Crowley AL, et al. Choice of agreement indices for
  assessing and improving measurement reproducibility in a core laboratory
  setting. {\it Statistical Methods in Medical Research} 2016\string;
  25(6)\string: 2939--2958.

\bibitem{jiang2017asymptotic}
Jiang J. {\it Asymptotic Analysis of Mixed Effects Models: Theory,
  Applications, and Open Problems}.
\newblock Boca Raton, Florida: CRC Press; 2017.

\bibitem{sas2017sas}
{SAS Institute Inc} . {\it SAS/STAT 9.4 User's Guide}.
\newblock Cary, NC: {SAS Institute Inc}; 2017.

\bibitem{fleiss1969large}
Fleiss JL, Cohen J, Everitt BS. Large sample standard errors of kappa and
  weighted kappa. {\it Psychological Bulletin} 1969\string; 72(5)\string:
  323--327.

\bibitem{psych19}
Revelle W. {\it psych: Procedures for Psychological, Psychometric, and
  Personality Research}. Northwestern University; Evanston, Illinois:  2018.
\newblock R package version 1.8.10.

\bibitem{oleckno2008epidemiology}
Oleckno WA. {\it Epidemiology: Concepts and Methods}.
\newblock Long Grove, IL: Waveland Press; 2008.

\bibitem{kuczmarska2016detection}
Kuczmarska A, Ngo LH, Guess J, et al. Detection of delirium in hospitalized
  older general medicine patients: A comparison of the {3D-CAM} and {CAM-ICU}.
  {\it Journal of General Internal Medicine}\string; 31(3)\string: 297--303.

\bibitem{vasunilashorn2016derivation}
Vasunilashorn SM, Guess J, Ngo L, et al. Derivation and validation of a
  severity scoring method for the 3-minute diagnostic interview for confusion
  assessment method--defined delirium. {\it Journal of the American Geriatrics
  Society} 2016\string; 64(8)\string: 1684--1689.

\bibitem{jiang2010large}
Jiang J. {\it Large Sample Techniques for Statistics}.
\newblock New York, NY: Springer Science \& Business Media; 2010.

\end{thebibliography}

\end{document}